%% file: main.tex
\documentclass{article}

\usepackage{arxiv}

\usepackage[utf8]{inputenc} 

\usepackage{amsfonts}       
\usepackage{nicefrac}       
\usepackage{microtype}      
\usepackage{lipsum}

\usepackage{color}
\usepackage{listings}
\usepackage{booktabs}
\usepackage{array}
\usepackage{graphicx}
\usepackage{longtable}
\usepackage{subfigure}

\usepackage{cite}
\usepackage{url}
\usepackage{fancyhdr}

\usepackage{soul}

\usepackage{float}
\usepackage{listings}
\usepackage{mdwmath}
\usepackage{mdwtab}
\usepackage{multirow}
\usepackage{multicol}
\usepackage{rotating}
\usepackage{setspace}
\usepackage[utf8]{inputenc}
\usepackage{lineno}
\usepackage{listings}

\usepackage{mdwmath}
\usepackage{mdwtab}
\usepackage{multirow}
\usepackage{multicol}
\usepackage{array}
\usepackage{booktabs}

\usepackage{enumitem}
\usepackage{xspace}
\usepackage[export]{adjustbox}
\usepackage{graphicx}
\usepackage{color,soul}
\usepackage{rotating}
\usepackage{setspace}
\usepackage{amsmath} 
\usepackage{amssymb}
\usepackage{float}
\usepackage{xcolor}

\usepackage{hyperref}
\usepackage{url}

\title{Decentralised Governance-Driven Architecture for Designing Foundation Model based Systems: Exploring the Role of Blockchain in Responsible AI}

\author{Yue Liu\textsuperscript{1,2}, Qinghua Lu\textsuperscript{1,2}, Liming Zhu\textsuperscript{1,2}, Hye-Young Paik\textsuperscript{2}\\
\textsuperscript{1}Data61, CSIRO, Australia\\
\textsuperscript{2}University of New South Wales, Australia
}

\begin{document}

\maketitle

\begin{abstract}
Foundation models including large language models (LLMs) are increasingly attracting interest worldwide for their distinguished capabilities and potential to perform a wide variety of tasks. Nevertheless, people are concerned about whether foundation model based AI systems are properly governed to ensure the trustworthiness and to prevent misuse that could harm humans, society and the environment. In this paper, we identify eight governance challenges of foundation model based AI systems regarding the three fundamental dimensions of governance: decision rights, incentives, and accountability. Furthermore, we explore the potential of blockchain as an architectural solution to address the challenges by providing a distributed ledger to facilitate decentralised governance. We present an architecture that demonstrates how blockchain can be leveraged to realise governance in foundation model based AI systems.

\end{abstract}

\textbf{Key terms - } Governance, Foundation model, Large language model, LLM, Blockchain, Accountability, \\ Responsible AI

\section{Introduction}

The year of 2023 has witnessed the emergence of large language models, one type of foundation models. Unlike previous AI models which are trained using data from a particular domain and aim to resolve problems in that domain, foundation models are trained with an extensive range of data for comprehensive capabilities, and eventually to accomplish various tasks~\cite{bommasani2022opportunities}. OpenAI released a conversational foundation model named ``ChatGPT" based on the GPT-3.5 model in 2022. ChatGPT draws widespread attention from diverse areas that it reached over 100 million users in two months after release~\cite{teubner2023welcome}, resulting in competition that many other IT companies also released their foundation models: Google introduced ``Bard", and Meta released ``LLaMA", to name a few.

\begin{figure}[tb]
  \centering
  \includegraphics[width=0.75\columnwidth]{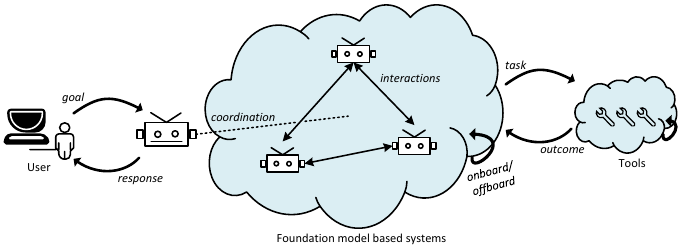}
  \caption{Human-AI teaming in job market.}
  \label{fig:ecosystem}
\end{figure}

Foundation models can provide a wide variety of services based on the massive AI models and vast amounts of broad training data~\cite{bommasani2022opportunities}. For instance, ChatGPT has shown its outperformance in natural language processing, code programming and analysis. Currently there are numerous projects exploring the potential use of foundation model based systems (e.g. agents) in diverse human-AI teaming scenarios, such as climate~\cite{LEIPPOLD2023103617}, medicine~\cite{kung2023performance}, gaming~\cite{park2023generative}, etc., Figure~\ref{fig:ecosystem} is a high-level graphical representation of foundation model based system ecosystem. Users can sift out the appropriate foundation model based system(s) considering the performance, cost, etc. to achieve certain goals. When multiple systems are employed, they may cooperate with each other while one of them needs to act as a coordinator. The systems can generate strategies and tasks, which may require subtle tools to orchestrate chores such as interactions and specific problem resolution.

However, we notice that currently there is a lack of consideration of how to realise governance in foundation model based AI systems across different architecture layers. Specifically, foundation model based systems consist of diverse AI and non-AI components, plugins, and stakeholders behind these products. Lacking proper governance solutions may result in disordered decision rights, complicated accountability processes, and eventually harm to humans, society and the environment. Considering the diversity of involved stakeholders, blockchain can serve as an infrastructure to realise decentralised governance in foundation model based AI systems.

Blockchain was firstly known as the underlying infrastructure of cryptocurrencies~\cite{Satoshi:bitcoin}. It is then generalised to distributed data storage and computing platforms, where a large network of untrusted participants need to reach agreements on transactional data states~\cite{scheuermann2015iacr}. Blockchain technology is proven to preserve certain software attributes and bring its distinctive features like transparency, on-chain autonomy to existing software applications. Blockchain can provide two core elements for realising decentralisation in existing software systems: (i) a distributed ledger, and (ii) a decentralised ``compute'' infrastructure. A blockchain is essentially a distributed ledger for transaction storage and verification without relying on any central trusted authority \cite{scheuermann2015iacr}. The on-chain programmability via smart contracts enables blockchain as a ``compute'' infrastructures~\cite{Omohundro:2014}. Smart contracts can be deployed on-chain to realise complex on-chain business logic such as triggers, conditions, etc.

In this paper, blockchain is leveraged as a means to realise governance in foundation model based systems. We first analyse the challenges in the foundation model systems regarding the three governance dimensions, namely, decision rights, incentives, and accountability. We then discuss how blockchain can address each identified challenge, and present a blockchain-based architecture design for governance-driven foundation model systems, where blockchain is leveraged for identity management, response recording and validation, and incentive distribution.

\section{Governance challenges in foundation model based systems}
\label{sec:Challenges}
Foundation models are believed to become a revolutionary force in the field of artificial intelligence, whereas the development and use of foundation model based AI systems are still at an early stage. It is in doubt whether these systems can behave in a responsible and trustworthy manner. In this paper, we adhered to our previous work on designing foundation model based systems~\cite{lu2023responsible}, and extended the scope by adopting traditional IT governance dimensions~\cite{ITgovernance} and exploring European Commission's new Product Liability Directive~\cite{New_Product_Liability_Directive, rodriguez2023revision} to investigate the governance issues. As listed in Table~\ref{tab:challenges}, there is still a set of governance challenges in foundation model based AI systems.

\begin{table*}[tbp]
\footnotesize
\centering
\caption{Governance challenges of foundation model based AI systems.}
\label{tab:challenges}
\begin{tabular}{p{0.1\textwidth}p{0.01\textwidth}p{0.35\textwidth}p{0.43\textwidth}}
\toprule

{\bf Governance dimensions} & &
\multirow{2}{0.35\textwidth}{\bf \centering  Challenges} &
\multirow{2}{0.4\textwidth}{\bf \centering  Blockchain-based solutions}
\\
\midrule

\multirow{11}{0.1\columnwidth}{\textbf{Decision rights}} 
& \multirow{4}{0.01\columnwidth}{D1.} & \multirow{4}{0.35\textwidth}{How to determine the decision rights of stakeholders, and how they can control and act in the system?} & 
Blockchain provides a governance infrastructure where stakeholders' decision rights can be managed via embedded access control mechanisms (e.g., who can access the training dataset for foundation models). \\
\cmidrule{2-4}

& \multirow{3}{0.01\columnwidth}{D2.} & \multirow{3}{0.35\textwidth}{How to determine the Intellectual Property (IP) of contents generated by foundation model based systems?} & 
IP agreement template can be deployed as smart contracts, which should be signed by the involved stakeholders (e.g. system providers, foundation model providers, users).\\
\cmidrule{2-4}

& \multirow{3}{0.01\columnwidth}{D3.} & \multirow{3}{0.35\textwidth}{How to select foundation models, agents, or external tools for certain tasks?} & 
A marketplace of foundation model based systems or external tools can be developed and deployed as a decentralised application in a blockchain network.\\
\cmidrule{1-4}

\multirow{8.5}{0.1\columnwidth}{\textbf{Incentives}}
& \multirow{5}{0.01\columnwidth}{I1.} & \multirow{5}{0.35\textwidth}{How to motivate the foundation model based systems to behave in a responsible manner?} & 
Blockchain can be used to distribute incentives to reward stakeholders (e.g., verifiers, tool providers) for actions that are aligned with human values, accordingly, incentives will also be locked or destructed if violations are detected.\\
\cmidrule{2-4}

& \multirow{3}{0.01\columnwidth}{I2.} & How to compensate stakeholders who are affected by the unintended or harmful behaviours of foundation model based systems? &
\multirow{3}{0.43\textwidth}{The impacted stakeholders can be registered in a smart contract. Compensation can be made after confirmation.} \\
\cmidrule{1-4}

\multirow{14}{0.1\columnwidth}{\textbf{Accountability}}
& \multirow{4}{0.01\columnwidth}{A1.} & \multirow{4}{0.35\textwidth}{How is identity managed in foundation model based AI systems?} &
Stakeholders can participate in the same blockchain network, where they can register blockchain accounts as on-chain identities for themselves and their products or virtual representatives (e.g., agents).\\
\cmidrule{2-4}

& \multirow{4}{0.01\columnwidth}{A2.} & \multirow{4}{0.35\textwidth}{How to scrutinise the operation information of foundation models?} &
Smart contracts can provide storage for recording users' prompts and foundation model-generated responses, and voting schemes for reaching consensus on the validation results. \\
\cmidrule{2-4}

& \multirow{6}{0.01\columnwidth}{A3.} & \multirow{6}{0.35\textwidth}{How to realise responsible resource provenance in foundation model based AI systems?} &

Blockchain can be leveraged to record critical runtime data of different components, such as inputs/outputs of foundation models, actions taken by the external tools, retrieved data from local data store through RAG, etc. Such information can enable traceability and auditability of the responsibilities of relevant stakeholders.\\

\bottomrule
\end{tabular}
\end{table*}

\textbf{Challenges of decision rights.} 
Decision rights refer to stakeholders' authority, responsibility and capability for decision-making. The determination of decision rights is complicated as there are diverse roles in foundation model based systems. For instance, a system consists of a foundation model, the orchestration components for handling interactions and communication, external systems and corresponding APIs and plugins for certain tasks, and additional operational components for safeguarding the use of foundation model. All components have their respective providers. A critical governance issue is that what are the responsibilities and capabilities of these project teams in a foundation model based system, and how they control the rights and act in the system to coordinate with users.

Further, for a foundation model based system, users input prompts to the system, and receive the generated responses. Nevertheless, the generated content may cause conflicts as users believe the results are created adhering to their thoughts and instructions, whilst the system providers may also require the responses to fine-tune the foundation models and train new models. Hence, the involved parties of a foundation model based system need to decide the rights of the model-generated content and related intellectual property. A more intricate circumstance is to take the data subjects and their consent for a foundation model to perform tasks into consideration.

In addition, there may be highly-modularised systems that allow users to customise the combination of multiple foundation model based systems and plugin tools, which will derive the need of comparing and selecting different systems and downstream tools for certain tasks. System developers, procurers, and users need to determine which system to deploy or even whether to include multiple systems to collaborate. A foundation model based system marketplace is required to provide a unified and convenient source to select assorted systems and tools based on particular metrics (e.g., price, processing time, context window).

\textbf{Challenges of incentives}. 
In a foundation model based system, any misconduct during operation may result in the abnormal performance of foundation model, e.g., mistakes in model fine-tuning can lead to incorrect responses to users' prompts. Incentives are considered a significant factor to motivate and regulate the system and related stakeholders' behaviours. We also envision that the systems themselves may require more rights in future, therefore, users may need to provide actual incentives to accomplish certain tasks. 

In addition, product defects may lead to the unintended or harmful behaviours of foundation model systems (e.g., property damage due to inaccurate classification outputs), which could heavily affect end users and even a larger community~\cite{New_Product_Liability_Directive}. In this case, the system or tool providers are liable for the defects and unexpected behaviours, while the victims may demand restitution. Hence, a complete process of compensation is required for foundation model based AI systems. Such situations can be further generalised to dispute resolution in foundation model systems. Different stakeholders may have their own goals, while the authorities in a foundation model based system (e.g., system providers, governors) can resolve conflicts by providing incentives to emphasise the collective benefit of most stakeholders. However, a subsequent concern is whether pursuing incentives introduces bias in prompt execution and conflict resolution, and corresponding mechanisms should be in place to redress issues or harms.

\textbf{Challenges of accountability}. 
Accountability refers to the identifiability, answerability and traceability processes of stakeholders for their behaviours and activities in corresponding decision-making processes. A fundamental challenge of accountability would be how to manage identities in foundation model based AI systems. A foundation model system may be developed and operated across multiple organisations, and a unified identity management solution, albeit each organisation may have its own management system, can help facilitate the coordination across different organisations. Further, with the increasingly enhanced capability of foundation models, the application systems may be assigned formal identities in future to better address accountability issues.

A foundation model is developed via a large amount of corpus for training and specific datasets for fine-tuning, and the procurers will adapt the foundation model and implement the entire AI system. Nevertheless, ineffective training and fine-tuning, tooling defects, or misleading instructions (e.g., prompt injection) may introduce discrimination to the system, and result in biased responses or hallucinations to users. Consequently, responses need to be examined to understand and assess the operation status and mitigate risks. Verifiers are incorporated to validate model-generated results for iterative fine-tuning, while their activities also need proper supervision to ensure security and safety.

The process of acquiring necessary data and tooling resources (e.g. external software systems) for accomplishing certain tasks can be sophisticated as it involves problems such as data privacy and security, obtaining access to specific tools or APIs, licensing agreements, etc. Maintaining proper recordings for responsible provenance process is significant for realising accountability of the eventual responses to users.

\begin{figure*}[htb]
  \centering
  \includegraphics[width=\textwidth]{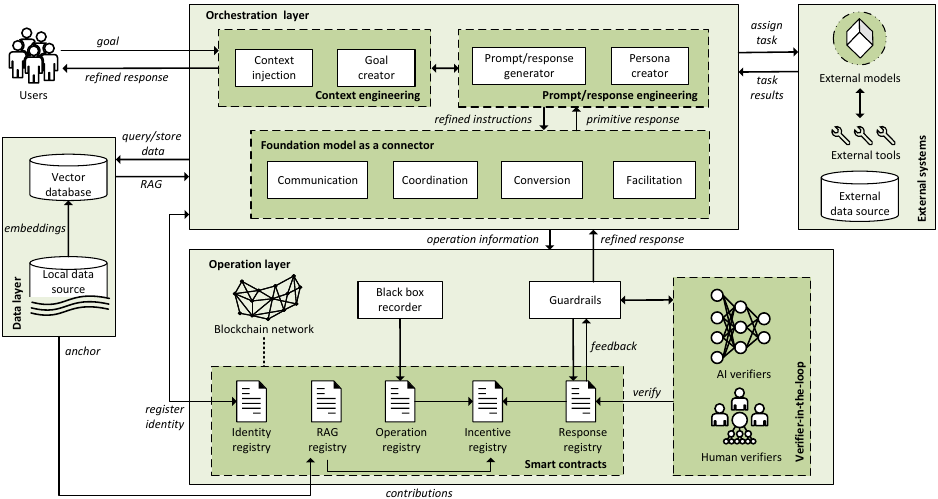}
  \caption{Blockchain-based architecture for governance-driven foundation model applications.}
  \label{fig:architecture}
\end{figure*}

\section{Blockchain as a Governance Solution for Foundation Model Systems.}
\label{sec:solution}

Blockchain can be leveraged to address governance issues by providing a transparent distributed ledger maintained in each participating entity for accountability, and a programmable infrastructure to facilitate and automate incentive distribution and decision-making processes. The third column of Table~\ref{tab:challenges} specifies how blockchain can be utilised to realise governance in foundation model based systems regarding the above identified challenges.

First, the system providers can deploy a permissioned blockchain network where all relevant stakeholders can collaborate to finalise different decisions. Specifically, smart contracts enable built-in access control to manage stakeholders' decision rights to certain governance issues (\textbf{\textit{D1}}). For instance, only appointed verifiers are capable of viewing and validating model-generated responses, or only system providers and auditors can access the training dataset. All participating stakeholders will have their public keys as on-chain identities and private keys for authorisation. Note that the whole blockchain network can be considered a mapping to off-chain business relationships, hence participating in the network may require real-world identity verification, which can ensure more complete accountability attribution (\textbf{\textit{A1}}). Further, blockchain-based self-sovereign identity can help establish formal decentralised identities for different users, foundation models, systems and tools in business relationships.

When operating a foundation model based system, the system providers can explain the intellectual property rights of model-generated data, and obtain users' consent to the clarified terms and policies, which can be demonstrated in the form of smart contracts. The system providers, users and all involved stakeholders need to reach agreement and sign the smart contract before users can access the system services (\textbf{\textit{D2}}). In addition, assorted systems can provide similar services, whilst a highly-modularised system may consist of multiple foundation models and tools, and users can choose one or several systems/models/tools for certain tasks. In this case, a marketplace for selecting foundation model systems can be developed as a decentralised application on blockchain, to provide a unified infrastructure for users to intuitively compare the systems regarding different metrics (\textbf{\textit{D3}}).

Blockchain allows issuing on-chain tokens\footnote{Programmable digital assets, different from the concept of ``token" as input contents in foundation models.} as inherent incentives to motivate stakeholders' behaviours. For instance, verifiers are rewarded for their contributions of improving foundation model's response. The system providers can either issue two types of tokens to represent the positive and negative incentives respectively, or manage only one token type for both rewards and penalties (e.g., locking or destructing tokens) via clear and strict regulations. Stakeholders' possessed tokens are checked periodically and transferred into real-world currencies based on the business agreement. Further, incentives can be given to the systems or foundation models as they also have on-chain identities, to encourage or depress certain activities (\textbf{\textit{I1}}). However, a concern would be that foundation models are considered ``black box" to stakeholders, it would be difficult to anticipate how they would perform to gain incentives, which may require enhanced explainability to improve trustworthiness.

A broad community can benefit from the services provided by the systems, or may be affected by their unintended or harmful behaviours. In the latter case, blockchain can serve as a log for identifying the responsible entities, and hence alleviating users' burden of proof. The victims of foundation model systems may request compensation for their loss. Smart contracts provide distributed registries where all impacted entities can sign up, while the system providers and other responsible stakeholders can make compensation according to these registries after accountability processes (\textbf{\textit{I2}}).

In particular, the foundation model system operation status can be revealed by the generated responses, which can be recorded in smart contracts for validation. Nevertheless, verifiers may have opposite understandings and thoughts about the responses. Such conflicts can be resolved via efficient ways such as the authorities (e.g., system providers) can make decisions in a short time, or via more democratic means like referendum. In the latter case, smart contracts support voting for validation determination, where various voting schemes can be applied to highlight different quality attributes (e.g., security, flexibility, preference expression) (\textbf{\textit{A2}}). System providers can also participate in voting as their suggestions are significant to verifiers, and their vote(s) can be set with a different weight. 

Meanwhile, all activities related to response validation are recorded by blockchain for further analysis. In addition, for each task, the critical runtime data and resource information (e.g., foundation model input/output, external tools, retrieved data from local data store through retrieval augmented generation (RAG)) can all be kept in smart contracts for auditing (\textbf{\textit{A3}}). All on-chain data cannot be tampered with or discarded by malicious stakeholders without being perceived by others. Consequently, blockchain can ensure data integrity, hence resource provenance and accountability in foundation model based AI systems by providing evidence for audit trail.

\section{Blockchain-based architecture for governance-driven foundation model based systems.}
\label{sec:architecture}

In this section, we propose an architecture design to present how to leverage blockchain as a software component to address certain governance issues in foundation model based AI systems. We adopt and extend existing architectures for foundation model based AI systems~\cite{lu2023responsible, lu2023taxonomy, lu2023towards, EmergingLLM} by integrating a blockchain network and five on-chain smart contracts. Figure~\ref{fig:architecture} illustrates an overview of a blockchain-based governance-driven architecture design for foundation model based systems. Specifically, the architecture consists of three main layers: orchestration layer, data layer, and operation layer, while we also include external application systems to demonstrate the complete workflow.

\textbf{Orchestration layer.} 
The orchestration layer maintains the core services of the whole foundation model system. It is responsible for receiving, executing, and replying to users' instructions, and connecting with other layers for task completion. In particular, the system provides \textit{context engineering} for handling users' inputs to understand the ultimate goals, which are then processed by \textit{prompt/response engineering} components, e.g., inspect whether there are prompt injections, modify or refuse the instructions, and refine the responses. Valid prompts are transferred to the foundation model, which can be leveraged as a \textit{software connector}~\cite{lu2023taxonomy} in the orchestration layer as follows:

\begin{itemize}
    \item \textit{Communication}: Foundation model transfers data between software components, e.g., sending certain data to external applications.

    \item \textit{Coordination}: Foundation model coordinates the computation results, e.g., decomposing an assignment into fine-grained tasks and generating execution plans.

    \item \textit{Conversion}: Foundation model transforms data format to assist communication between components, e.g., interpreting users' descriptions to other AI models in a machine-readable scheme.

    \item \textit{Facilitation}: Foundation model optimises the overall workflow and interactions between components by finalising specific decisions, e.g., whether to invoke other components.
    
\end{itemize}

Meanwhile, completing tasks may require external systems, which include other AI models and tools for specific computation processes, and additional data sources for useful information. The task/search results are sent back to the orchestration layer for further processing by foundation model, and storage in the data layer for future queries.

\textbf{Data layer.}
The data layer includes two main components: \textit{local data source} (e.g., data lake) and \textit{vector database}. The local data source is responsible for storing raw data, which then will be converted into embeddings (i.e., numerical representations of data) and recorded in the vector database. Specifically, the embeddings can represent the semantic meaning of data, enabling efficient similarity search. As there is usually a context window in foundation models, applying a vector database can help achieve better accuracy for foundation models to understand users' input and generate responses through retrieval augmented generation.

\textbf{Operation layer.} 
The operation layer consists of a set of on-chain smart contracts and the related off-chain components: \textit{black box recorder}, \textit{guardrails}, \textit{verifier-in-the-loop}, a \textit{blockchain network} and on-chain smart contracts: \textit{identity registry, RAG registry, operation registry, response registry}, and \textit{incentive registry}. 

First, blockchain provides a decentralised public key infrastructure that can assign blockchain accounts to all stakeholders. Moreover, the \textit{identity registry} can enable formal on-chain identity management (e.g., self-sovereign identity) to establish and maintain certain business relationships. 

Secondly, considering off-chain data repositories may be compromised and tampered with, which will result in inaccurate model responses, the \textit{RAG registry} can be utilised for periodically anchoring off-chain data from the data layer. Similarly, \textit{black box recorder} saves the runtime data in \textit{operation registry}, including the input/output, and intermediate data of other layers and components (e.g., external tools). In addition, user consent can also be included in the \textit{operation registry} as it is regarded as the starting point for a task. Storing all the above information on-chain comprises a complete and traceable workflow to provide audit trails if the system has unexpected behaviours. Any attempt to revise on-chain stored data will leave traces while modifying the related block requires altering all subsequent blocks, hence data integrity is preserved via smart contracts.

Thirdly, \textit{response registry} facilitates the \textit{guardrail} functionalities. Guardrails can verify whether users' prompts are compliant with responsible AI requirements and also refine foundation model's responses. Any response violating system provider's predefined rules will be recorded in \textit{response registry}. Subsequently, verifiers can assess the quality (e.g., correctness, relevance, and appropriateness) of model outputs and guardrails operation via this registry and provide feedback (e.g., voting whether the output is appropriate and useful), which is then used to improve the guardrails and fine-tune the foundation model to better process users' instructions.

Finally, the \textit{incentive registry} records the contributions of different stakeholders. In particular, \textit{RAG registry} and \textit{operation registry} transfer data contributors and providers of the employed tools respectively, while \textit{response registry} delivers the involved verifiers to \textit{incentive registry}. System providers can either issue and distribute on-chain tokens via this registry, or examine the contribution records and provide rewards to the relevant entities through off-chain channels.

\textbf{Discussion.} 
In the proposed architecture, blockchain is exploited to achieve governance in foundation model based systems. First, blockchain assists in decision-making on response validation via various voting schemes, which can also be applied to other conflict resolutions in the system. Moreover, the providers of foundation model and external applications can register identities for their products to enable accountability and incentive distribution processes. Responsibility attribution is facilitated when the foundation model generates incorrect or even malicious responses to users, since blockchain records the used data, runtime information, and prompt execution outcome, along with the registered on-chain identities of involved stakeholders. In addition, incentives are allocated according to the contributions regarding data, tooling, task completion and validation.

We remark that the proposed architecture can serve as a reference while the practitioners need to finalise the design decisions further. For instance, a consortium blockchain network can be deployed with resource/energy-efficient solutions. The prompts and responses can be encrypted or apply certain access control mechanisms to be examined by only authorised verifiers. The votes can be held with different schemes, such as one verifier one vote, one token one vote, etc. It should also be noted that the incentive distribution to foundation models is dependent on model explainability, whilst the marketplace for foundation model systems also requires clear taxonomy based on assorted metrics.

\section{Conclusion and Future Work}
\label{sec:conclusion}

In this paper, we identify a series of governance challenges in foundation model based AI systems in terms of decision rights, incentives, and accountability, and discuss the role of blockchain in addressing each identified challenge. In addition, we propose a blockchain-based architecture for designing governance-driven foundation model systems. The architecture focuses on the governance issues during the operation of foundation model systems, including identity management, response validation, incentive provision, and accountability of system operation. The proposed architecture leverages five on-chain smart contracts to realise governance in the foundation model based system. In our future work, we plan to implement a proof-of-concept prototype with fine-grained design decisions, evaluate its performance, and further explore decentralised governance in foundation model based systems.

\input{main.bbl}


\end{document}